\documentclass[aps, prx, preprint, superscriptaddress, showkeys]{revtex4-2}
\usepackage{graphicx}

\begin{document}

\renewcommand{\figurename}{FIG.}

\title{Time-resolved measurement of ambipolar edge magnetoplasmon transport in InAs/InGaSb composite quantum wells}


\author{H. Kamata}
\affiliation{NTT Basic Research Laboratories, NTT Corporation, 3-1 Morinosato-Wakamiya, Atsugi, Kanagawa 243-0198, Japan}
\affiliation{JST, PRESTO, 4-1-8 Honcho, Kawaguchi, Saitama 332-0012, Japan}
\author{H. Irie}
\affiliation{NTT Basic Research Laboratories, NTT Corporation, 3-1 Morinosato-Wakamiya, Atsugi, Kanagawa 243-0198, Japan}
\author{N. Kumada}
\affiliation{NTT Basic Research Laboratories, NTT Corporation, 3-1 Morinosato-Wakamiya, Atsugi, Kanagawa 243-0198, Japan}
\author{K. Muraki}
\affiliation{NTT Basic Research Laboratories, NTT Corporation, 3-1 Morinosato-Wakamiya, Atsugi, Kanagawa 243-0198, Japan}

\date{\today}%

\begin{abstract}
Time-resolved charge transport measurement for one-dimensional edge states is a powerful means for investigating nonequilibrium charge dynamics and underlying interaction effects therein.
Here, we report a versatile on-chip time-resolved transport measurement scheme that does not require a quantum point contact and is therefore applicable to narrow-gap systems.
We apply the technique to non-inverted InAs/In$_{x}$Ga$_{1-x}$Sb composite quantum wells, where its ambipolar character enables us to demonstrate the scheme in both the electron and hole regimes separately using a single device.
Time-resolved measurements in the quantum Hall regimes clearly exhibit the chirality of each carrier, with pulsed charge waveforms observed only for one magnetic field direction opposite for electrons and holes.
Waveform analysis in the time domain reveals reduced group velocity and broadening of edge magnetoplasmon pulses in both the electron and hole regimes, suggesting the influence of charge puddles in the bulk.
Our time-resolved measurement scheme, applicable to various systems, will pave the way for investigations of dynamical properties of exotic topological edge states.
\end{abstract}

\maketitle

\section{Introduction}
\label{Intro}

Two-dimensional (2D) topological insulators have attracted interest for their unique properties of gapless edge states, which dominate the transport and low-energy excitation of the system~\cite{hasan2010colloquium}.
While edge states have been studied mainly by low-frequency or dc transport measurements, high-frequency measurements provide information about dynamics not attainable otherwise.
Frequency- or time-domain measurements performed on chiral edge states of quantum Hall (QH) systems, the prototypical 2D topological insulator with broken time-reversal symmetry, have revealed the dynamics of nonequilibrium charge propagating unidirectionally as edge magnetoplasmon (EMP) modes~\cite{ashoori1992edge, zhitenev1994experimental, ernst1997dynamic, sukhodub2004observation, kamata2010voltage, kumada2011edge, kumada2013plasmon, kumada2020suppression}, elucidating the underlying Coulomb interaction and the Tomonaga-Luttinger liquid behavior~\cite{berg2009fractional, bocquillon2013separation, kamata2014fractionalized, freulon2015hong, marguerite2016decoherence, hashisaka2017waveform, brasseur2017charge, hashisaka2018tomonaga}.
It is therefore interesting to explore spin/charge dynamics and underlying interaction effects in helical edge states of quantum spin Hall (QSH) systems, the time-reversal-invariant counterpart of the QH system, where up and down spins move in opposite directions.

Despite various proposals~\cite{hofer2014proposal, calzona2015time, dolcini2016photoexcitation, muller2017dynamical, acciai2019spectral, gourmelon2020characterization}, dynamics in helical edge states remain largely unexplored experimentally.
While microwave spectroscopy has recently been used to disentangle edge and bulk transport in HgTe/(Hg)CdTe quantum wells in the frequency domain~\cite{dartiailh2020dynamical}, time-domain waveform measurements of  edge plasmon modes remain a challenge.
One of the major obstacles is the narrow band gap of systems that support the QSH phase such as HgTe/(Hg)CdTe~\cite{bernevig2006quantum, konig2007quantum} and InAs/(In)GaSb~\cite{liu2008quantum, knez2011evidence, du2015robust, akiho2016engineering, couedo2016single, du2017tuning, irie2020energy} quantum wells.
While, in QH systems, direct measurements of EMP transport have been achieved using a sampling oscilloscope~\cite{kumada2013plasmon, brasseur2017charge}, this technique requires a large rf excitation, which could induce undesired bulk mode excitations in the topological regime of narrow-gap systems.
An on-chip time-resolved charge detection scheme using a quantum point contact (QPC) as a detector provides a powerful means of measuring charge waveforms with higher sensitivity~\cite{kamata2010voltage, kamata2014fractionalized, kataoka2016time, hashisaka2017waveform, roussely2018unveiling, lin2021quantized, lin2021time} and thus can operate at lower rf excitation.
However, as the scheme relies on the temporal modulation of the QPC opening near the pinch-off~\cite{kamata2009correlation}, it is not straightforward to implement it in zero- or narrow-gap systems where charge carriers cannot be completely depleted.
It is therefore essential to develop an alternative measurement technique for high-frequency charge transport applicable to narrow-gap systems.

In this paper, we report an on-chip time-resolved transport measurement scheme for narrow-gap systems based on a temporal modulation of reflection/transmission through a narrow constriction.
We demonstrate this scheme through measurements of noninverted InAs/In$_{x}$Ga$_{1-x}$Sb composite  quantum  wells (CQWs)~\cite{irie2020energy} in the QH regime.
The ambipolar character of the InAs/In$_{x}$Ga$_{1-x}$Sb CQWs enables us to investigate nonequilibrium charge dynamics in both electron and hole regimes separately using a single device.
The data clearly exhibit the chirality of the charge carriers, with pulsed charge waveforms appearing only for one magnetic ﬁeld direction opposite for electrons and holes.
Waveform analysis in the time domain enables us to discuss dissipative transport of EMP modes in both regimes.
As the QSH phase can be approximately thought of as a superposition of QH chiral edge states of electron-like and hole-like character~\cite{buhmann2011the, young2014tunable}, our measurements for both electrons and holes in the QH regime represent an important step toward waveform measurements in the QSH regime.

This paper is organized as follows.
Section II describes the device and experimental method.
Section III presents results of on-chip time-resolved pulsed charge detection (IIIA), time-of-flight measurements in electron and hole regimes (IIIB), and a discussion based on waveform analysis (IIIC).
Section IV summarizes this paper.

\section{Experiment}
\subsection{Device structure and dc measurement}
\label{SampleDC}

The sample we studied is an InAs/In$_{\rm 0.25}$Ga$_{\rm 0.75}$Sb CQW grown on a semi-insulating GaAs substrate by molecular beam epitaxy.
Magnetotransport characterization of the sample shows that the CQW has a non-inverted band structure, with a single carrier type, either electron or hole, depending on the gate voltage (for details, see Supplemental Material).
The CQW comprises InAs (top) and In$_{\rm 0.25}$Ga$_{\rm 0.75}$Sb (bottom) wells with nominal thicknesses of 10 and 6~nm, which host 2D electrons and holes, respectively, for gate voltages above and below the charge neutrality point (CNP).
The CQW is sandwiched between Al$_{\rm 0.80}$Ga$_{\rm 0.20}$Sb barrier layers and located 55~nm below the surface.
The electron and hole mobilities measured in each regime at a carrier density of $2.0 \times 10^{15}$~m$^{-2}$ are 0.50 and 0.61~m$^{2}$/Vs, respectively.

\begin{figure}[ptb]
\begin{center}
\includegraphics[scale=1]{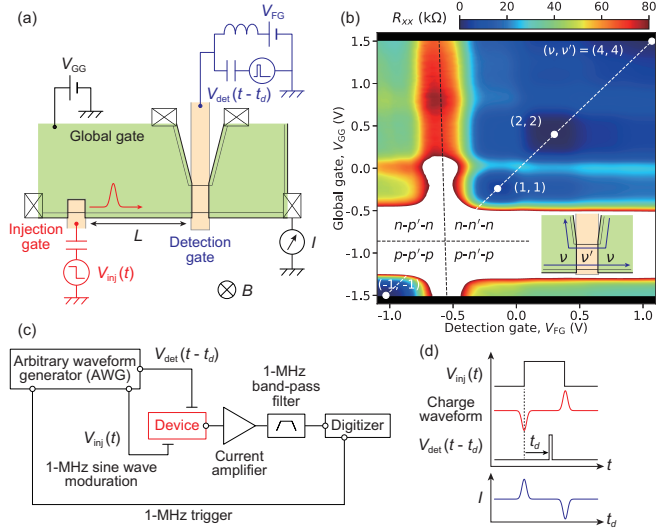}
\caption{
(a)
Schematic illustration of the device structure and experimental setup.
(b)
Color-scale plot of the longitudinal resistance $R_{xx}$ across the constriction at $B = 10$~T as a function of $V_{\rm FG}$ and $V_{\rm GG}$.
Electron and hole filling factors are labeled with positive and negative values, respectively.
The white area around the CNPs indicates the region where $R_{xx} > 80$~k$\Omega$.
The inset shows a schematic view of edge channels around the constriction in the QH regimes at $\nu = \nu^{\prime}$ (shown by white dots in the main plot).
(c)
Circuit diagram for the high-frequency measurement.
(d)
Schematic of charge waveform and voltage pulses $V_{\rm inj} (t)$ and $V_{\rm det} (t - t_{d})$ as a function of time $t$ (upper panel) with expected $I (t_{d})$ trace (lower panel).
The waveform of the excited charge pulse becomes the derivative of $V_{\rm inj} (t)$ through the capacitive coupling between the gate and the edge channels.
}
\label{Fig1}
\end{center}
\end{figure}

Figure~\ref{Fig1}(a) schematically shows the device structure and experimental setup for an on-chip time-resolved pulsed charge detection.
The device has two kinds of fine gates---one for exciting non-equilibrium charges (injection gate) and the other for filtering the propagating pulsed charges (detection gate)---and a global gate for tuning the density in the bulk through which the pulsed charges propagate.
The injection gate has a $3 \times 3$~$\mu$m$^{2}$ overlap with the lower edge of the mesa.
At a distance $L = 30$~$\mu$m to the right of the injection gate, the mesa has a narrow constriction, 5-$\mu$m long and 3-$\mu$m wide, which the detection gate covers.
The rest of the mesa regions are covered by the global gate.
The carrier types and the static carrier densities in the regions underneath the detection and global gates are controlled by dc gate voltages $V_{\rm FG}$ and $V_{\rm GG}$, respectively. For details of the device fabrication, see Appendix A.

Figure~\ref{Fig1}(b) shows the dc transport characteristics of the device, where we plot longitudinal resistance $R_{xx}$ across the constriction measured at a magnetic field $B = 10$~T as a function of $V_{\rm FG}$ and $V_{\rm GG}$.
The checkerboard pattern reflects the changes in the filling factors $\nu$ outside the constriction and $\nu^{\prime}$ inside it.
The horizontal and vertical dashed black lines represent the CNPs where $\nu = 0$ and $\nu^{\prime} = 0$, respectively.
These lines define four quadrants, $n$-$n^{\prime}$-$n$, $p$-$n^{\prime}$-$p$, $p$-$p^{\prime}$-$p$, and $n$-$p^{\prime}$-$n$, for the charge-carrier configurations across the constriction.
The diagonal dashed white line represents the condition $\nu = \nu^{\prime}$, where the carrier type and density become uniform across the constriction.
In the QH regimes with $\nu = \nu^{\prime}$, shown by white dots in Fig.~\ref{Fig1}(b), edge channels pass through the constriction as schematically shown in the inset of Fig.~\ref{Fig1}(b).

\subsection{Time-resolved measurement scheme}
\label{MeasScheme}

For time-resolved measurements, high-speed voltage pulses $V_{\rm inj} (t)$ and  $V_{\rm det} (t - t_{d})$, generated by an arbitrary waveform generator (AWG),  were applied to the injection and detection gates, respectively, with the rising and falling times of $\sim 100$~ps on the gates at a repetition frequency of 25~MHz [Fig.~\ref{Fig1}(c)].
The 20-ns wide injection pulse $V_{\rm inj} (t)$ excites negative and positive pulsed charges at its rising and falling edges, respectively [upper panel in Fig.~\ref{Fig1}(d)], which in the QH regime propagate along the mesa edge to the right or left, depending on the chirality determined by the field direction and the charge-carrier polarity.
The 80-ps wide detection pulse $V_{\rm det} (t - t_{d})$, much narrower than the injection pulse, has a time delay $t_{d}$ from the injection pulse and temporarily modulates the transmission of the pulsed charges through the constriction.
By measuring the current $I$ flowing out from the Ohmic contact located to the right of the constriction as a function of $t_{d}$, we obtain the charge waveform. When the transmission through the constriction is decreased by the detection pulse, the polarity of $I$ is reversed with respect to the pulsed charges [lower panel in Fig.~\ref{Fig1}(d)].
The time origin $t_{d} = 0$ was set at the timing when the rising edge of $V_{\rm det} (t - t_{d})$ coincides with that of $V_{\rm inj} (t)$ on the device.

In actual measurements, we further devised our scheme to obtain a reasonable signal-to-noise ratio.
First, the injection pulse $V_{\rm inj} (t)$ was modulated with a 1-MHz sine wave to eliminate undesirable direct high-frequency crosstalk between the detection pulse and the Ohmic contact as well as pulsed charges excited at the detection gate.
Then, the amplified current $I (t_{d})$ was sampled on a digitizer triggered by the AWG [Fig.~\ref{Fig1}(c)].
Finally, the recorded data were numerically averaged and then demodulated with a 1-MHz sine wave to obtain the waveform of interest.
We note that, while the dynamics we are looking at are in the gigahertz regime, our measurement scheme operates in the megahertz range.
This allows us to alleviate the issues common to direct high-frequency measurements in the gigahertz range, such as the dissipation of high-frequency signals on the Ohmic contact.
For details of the RF measurement sequence, see Appendix B.

All measurements were carried out at a temperature of 1.5~K in a magnetic field $B$ up to 11.5~T applied perpendicular to the sample.
For $B > 0$, the edge modes propagate counterclockwise (clockwise) in the electron (hole) regime.
Unless otherwise noted, the amplitudes of the injection and detection pulses are 100 and 200~mV$_{\rm pp}$, respectively.

\section{Results and discussion}
\subsection{Time-resolved pulsed charge detection mechanism}
\label{OnChip}

\begin{figure}[ptb]
\begin{center}
\includegraphics[scale=1]{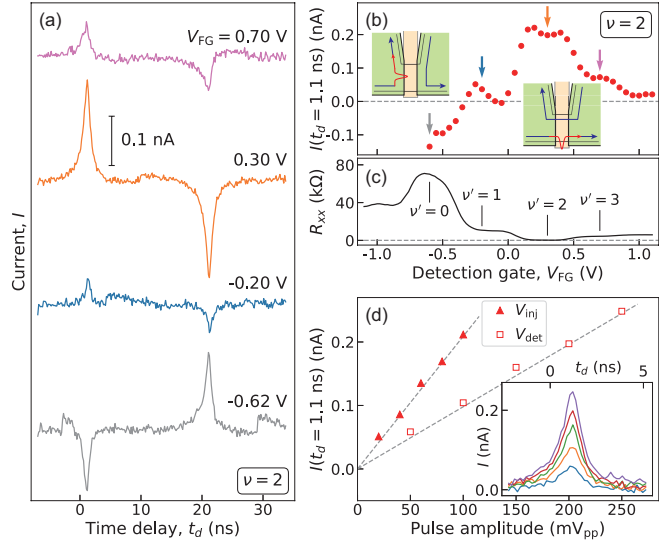}
\caption{
(a)
$I (t_{d})$ traces measured at $\nu = 2$ ($B = 10$~T, $V_{\rm GG} = 0.40$~V) for various $V_{\rm FG}$.
For clarity, the traces are vertically offset.
Abrupt jumps in $I(t_{d})$ seen in some traces are due to switching noise, which which occasionally appears around the CNP, with no reproducibility.
(b) $I(t_{d} = 1.1$~ns) and (c) $R_{xx}$ as a function of $V_{\rm FG}$.
Arrows represent the conditions where $\nu^{\prime}$ takes integer values.
The insets show schematic of pulsed charges around the constriction at $\nu^{\prime} = 0$ (left) and $\nu^{\prime} = \nu$ (right).
(d)
$I(t_{d} = 1.1$~ns) as a function of the pulse amplitudes of $V_{\rm inj}$ and $V_{\rm det}$ measured at $\nu^{\prime} = \nu = 2$ ($B = 10$~T, $V_{\rm FG} = 0.30$~V, $V_{\rm GG} = 0.40$~V).
The $V_{\rm det}$ ($V_{\rm inj}$) dependence was measured with the $V_{\rm inj}$ ($V_{\rm det}$) amplitude of 100 (200)~mV$_{\rm pp}$.
The inset shows $I(t_{d})$ traces around $t_{d} = 1.1$~ns for $50 \le V_{\rm det} \le 250$~mV$_{\rm pp}$ at $V_{\rm inj} = 100$~mV$_{\rm pp}$.
}
\label{Fig2}
\end{center}
\end{figure}

To see how our pulsed charge detection scheme works, we first compare charge waveforms measured at different $V_{\rm FG}$, i.e., at different filling factors $\nu^{\prime}$ in the constriction, with the filling factor $\nu$ in the bulk fixed.
Figure~\ref{Fig2}(a) shows the current $I$ as a function of the time delay $t_{d}$ for various $V_{\rm FG}$ (corresponding to $\nu^{\prime} = 3$, 2, 1, 0), measured at $\nu = 2$ ($B = 10$~T, $V_{\rm GG} = 0.40$~V), where the background current (typically $< 10$~nA depending on $\nu$ and $V_{\rm FG}$) is subtracted from each trace.
Each trace shows a pair of positive (peak) and negative (dip) signals at around $t_{d} = 1.1$ and 21.1~ns, which correspond to the charge pulses generated at the rising and falling edges of the injection pulse, respectively.
As $V_{\rm FG}$ is varied, not only the amplitude but also the polarity of the signal changes (see the trace for $V_{\rm FG} = -0.62$~V).
That is, depending on $V_{\rm FG}$, the measured current can increase or decrease with respect to the background level.

To examine the $V_{\rm FG}$ dependence in more detail, we plot in Fig.~\ref{Fig2}(b) the current $I$ at $t_{d} = 1.1$~ns as a function of $V_{\rm FG}$.
As a reference, we also plot in Fig.~\ref{Fig2}(c) $R_{xx}$ across the constriction separately measured at $\nu = 2$ without voltage pulses applied.
When $\nu^{\prime}$ is close to an integer, $R_{xx}$ exhibits a plateau at a value determined by $\nu^{\prime}$ and $\nu$ ($= 2$)~\cite{calvo2017interplay}.
The data reveal that $I(t_{d} = 1.1$~ns) oscillates with $V_{\rm FG}$, with local maxima at integer $\nu^{\prime}$, as indicated by arrows, and the global maximum at $\nu^{\prime} = 2$.
The data also reveal that the sign of the signal changes for $V_{\rm FG} < -0.3$~V.
We emphasize that the results essentially do not depend on the amplitudes of $V_{\rm inj} (t)$ and $V_{\rm det} (t - t_{d})$ in the relevant range.
As shown in Fig.~\ref{Fig2}(d), the signal amplitude measured at $V_{\rm FG} = 0.30$~V ($\nu^{\prime} = 2$) linearly increases with the amplitudes of $V_{\rm inj} (t)$ and $V_{\rm det} (t - t_{d})$ while maintaining the waveform [inset of Fig.~\ref{Fig2}(d)], indicating the linearity of the injector and detector.
In the following, we focus on the signal at $t_{d} \sim 1.1$~ns since the one at $t_{d} \sim 21.1$~ns has the same waveform (with reverse polarity).

The data shown in Fig.~\ref{Fig2} call for some discussion on the charge-detection mechanism in our scheme.
In the conventional on-chip time-resolved measurement scheme using a QPC as a charge detector~\cite{kamata2009correlation, kamata2010voltage}, the sensitivity is proportional to the change in the conductance through the QPC due to the detection pulse.
However, the $V_{\rm FG}$ dependence of the signal amplitude shown in Fig.~\ref{Fig2}(b) is obviously not consistent with this trend.
As opposed to the simple expectation that the sensitivity is proportional to $d R_{xx} / d V_{\rm FG}$, the signal amplitude becomes maximum in the QH regime where $d R_{xx} / d V_{\rm FG}$ is minimum.
As we discuss below, the observed $V_{\rm FG}$ dependence can be qualitatively understood by considering the reflection of pulsed charges at the entrance and exit of the constriction.
At $\nu^{\prime} = \nu$ ($= 2$), where the electrostatic environment inside and outside the constriction is nearly uniform, the pulsed charges pass through the constriction without reflection [right inset of Fig. 2(b)].
That is, reflection occurs only when the detection pulse $V_{\rm det} (t - t_{d})$ coincides with the arrival of the pulsed charges at the junction.
As a result, the measured signal amplitude reflects the degree of reflection of pulsed charges.
In contrast, at $\nu^{\prime} \neq \nu$ (e.g., $\nu^{\prime} = 1$, 3), the electrostatic environment and hence $\sigma_{xy}$ are spatially varied, and so the pulsed charges are partially reflected even without the detection pulse.
Consequently, the measured signal amplitude becomes smaller.
On the other hand, at $\nu^{\prime} = 0$ corresponding to the CNP, there is no chiral edge channel inside the constriction (i.e., $\sigma_{xy} = 0$), and, hence, almost all pulsed charges are reflected [left inset of Fig. 2(b)].
In this configuration, the detection pulse temporary increases the transmission through the junction.
This situation is similar to the conventional time-resolved measurement scheme with a relatively open QPC \cite{kamata2009correlation, kamata2010voltage}.
As a result, the charge polarity of waveforms measured in this regime is reversed.
In the following, all time-resolved measurements presented were carried out under the detection condition $\nu^{\prime} = \nu$, where the signal amplitude takes a maximum value.

\begin{figure}[t]
\begin{center}
\includegraphics[scale=1]{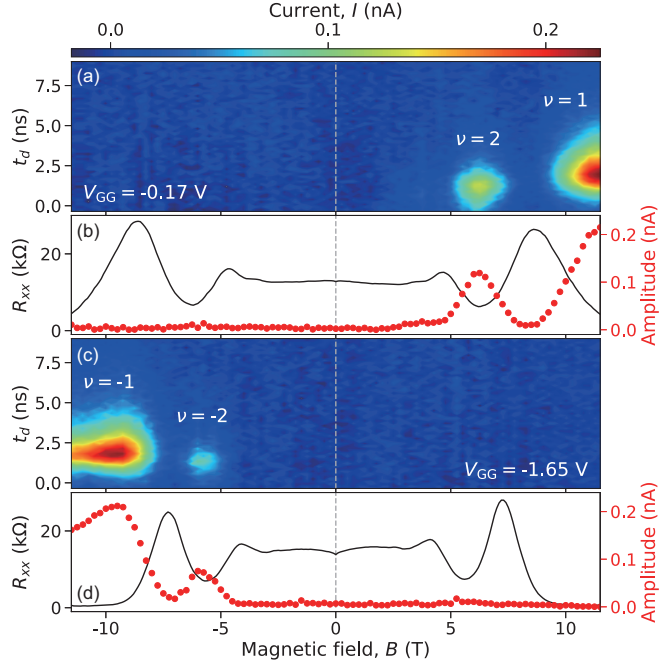}
\caption{
(a)
Color-scale plots of $I$ as a function of $t_{d}$ and $B$ for $V_{\rm GG} = -0.17$~V (electron regime).
(b)
$R_{xx}$ (black line) and signal amplitude (red dots) as a function of $B$ for $V_{\rm GG} = -0.17$~V (electron regime).
Panels (c) and (d) represent the same plots for $V_{\rm GG} = -1.65$~V (hole regime).
}
\label{Fig3}
\end{center}
\end{figure}

\subsection{Time-resolved chiral edge transport in the electron and hole regimes}
\label{Chirality}

Having established the charge pulse detection scheme for QH states in CQWs, we now use the global gate to investigate the EMP transport in the hole regime as well as the electron regime.
Figures~\ref{Fig3}(a) and \ref{Fig3}(c) show color-scale plots of $I$ as a function of $t_{d}$ and $B$, measured in the electron ($V_{\rm GG} = -0.17$~V) and hole ($V_{\rm GG} = -1.65$~V) regimes, respectively.
An important observation here is the symmetry between the waveforms measured in the electron and hole regimes. 
Namely, EMP signals appear only for one field direction that depends on the carrier polarity, $B > 0$ for electrons and $B < 0$ for holes.
The signal amplitude takes maximum values in the QH regime
[Figs.~\ref{Fig3}(b) and \ref{Fig3}(d)].
In both regimes, the peak shifts to larger $t_{d}$ as $|B|$ increases (and hence $|\nu|$ decreases) and, at the same time, broadens along the time axis.
The time position of the current peak roughly corresponds to the time-of-flight of EMP pulses for the path length ($L = 30$~$\mu$m).
As the signal amplitude significantly drops in the transition regions between QH states, in the following analysis we focus on the data taken at integer $\nu$.

\subsection{Waveform analysis}
\label{WaveAnalysis}

\begin{figure}[ptb]
\begin{center}
\includegraphics[scale=1]{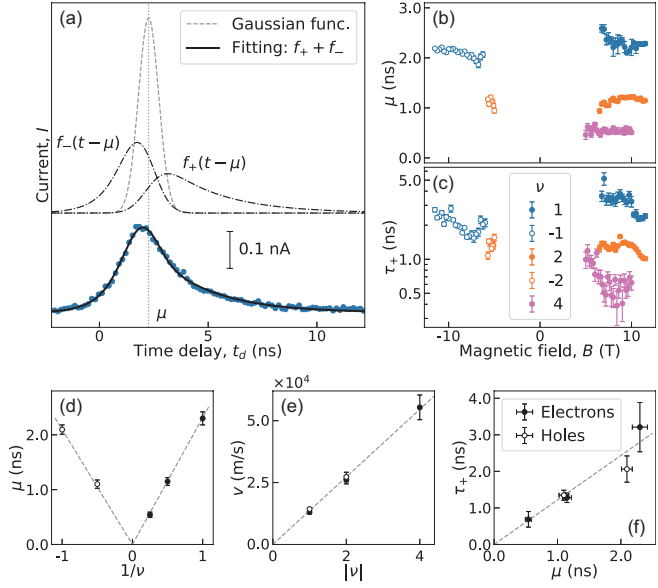}
\caption{
(a)
(Lower section) Waveform for $\nu = 1$ measured at $B = 10$~T (dots) with the fitting curve $f_{+}(t - \mu) + f_{-}(t - \mu)$ (solid lines).
(Upper section) Gaussian functions with (dashed-dotted) and without (dashed) exponential modification for the parameters used for $\nu = 1$.
The vertical dotted line represents the center position $\mu$ of the unmodified Gaussian function.
(b), (c)
$B$ dependence of
(b)
$\mu$ and
(c)
$\tau_{+}$ at each $\nu$.
(d)
$\mu$ as a function of $1 / \nu$ in the electron (solid circles) and hole (open circles) regimes.
The data are the averages for measurements at different $B$'s shown in (b).
The dashed lines are linear fits.
(e)
Velocity as a function of $|\nu|$ for the electron (solid circles) and hole (open circles) regimes with a linear fit.
(f)
$\tau_{+}$ as a function of $\mu$ for the electron (solid circles) and hole (open circles) regimes with a linear fit.
}
\label{Fig4}
\end{center}
\end{figure}

We analyze the measured waveforms in more detail and deduce the mode velocity and broadening of the EMP pulse, which provide information about the electrostatic environment around edge channels~\cite{kumada2011edge, kumada2020suppression, lin2021time}.
In Fig.~\ref{Fig4}(a), we plot the $I(t_{d})$ trace for $\nu = 1$ (dots) taken at $B = 10$~T.
The waveform is asymmetric, with a longer tail on the right side of the peak.
We evaluate the asymmetric waveform by fitting with an exponentially modified Gaussian function given by
\begin{widetext}
\begin{equation}
\label{exGauss}
f_{\pm}(x)
= \frac{a \sigma}{\tau_{\pm}}
\sqrt{\frac{\pi}{2}}
\exp
\left(
\frac{1}{2}
\left(
\frac{\sigma}{\tau_{\pm}}
\right)^{2}
\mp \frac{x - \mu}{\tau_{\pm}}
\right) \
{\rm erfc}
\left(
\frac{1}{\sqrt{2}}
\left(
\frac{\sigma}{\tau_{\pm}}
\mp \frac{x - \mu}{\sigma}
\right)
\right),
\end{equation}
\begin{equation}
\label{erfc}
{\rm erfc} (x) = 1 – {\rm erf} (x)
= \frac{2}{\sqrt{\pi}}
\int^{\infty}_{x} e^{-t^{2}} dt,
\end{equation}
\end{widetext}
where $\tau_{+}$ ($\tau_{-}$) is the exponent relaxation time giving a positive (negative) skew to the Gaussian function, $a$,  $\mu$, and $\sigma^{2}$ are the amplitude, center position and variance of the Gaussian function, respectively, and Eq.~(\ref{erfc}) is the complementary error function~\cite{purushothaman2017hyper}.
In our on-chip detection scheme, the measured $I(t_{d})$ trace has the form of a convolution between the EMP pulse waveform at the detector and the time-dependent local potential there modulated by the detection pulse.
Accordingly, $\tau_{+}$ and $\tau_{-}$ predominately reflect the broadening of the EMP pulse and the response time of the local potential, respectively~\cite{kamata2009correlation}.

As shown in Fig.~\ref{Fig4}(a), a superposition of $f_{+}(t - \mu)$ and $f_{-}(t - \mu)$ (solid lines) well reproduces the measured waveform.
The individual components of $f_{+}(t - \mu)$ and $f_{-}(t - \mu)$ used to fit the waveform are shown (dashed-dotted lines) together with the unmodified Gaussian function with the same $a$, $\mu$, and $\sigma$ (dashed line).
Note that it is $\mu$, and not the peak position of $I(t_{d})$, that corresponds to the actual time-of-flight of EMP pulses when the dispersion is linear.
This is important as the peak position of $I(t_{d})$ does not coincide with that of the Gaussian function when $\tau_{+} \neq \tau_{-}$ as is the case here.

We performed similar measurements at different magnetic fields for electrons ($\nu = 1$, 2, and 4) and holes ($\nu = -1$ and $-2$) by tuning $V_{\rm GG}$ to maintain $\nu$ constant (see Supplemental Material).
The analysis results for these measurements are summarized in Figs.~\ref{Fig4}(b) and \ref{Fig4}(c), where we plot $\mu$ and $\tau_{+}$, respectively, obtained for $\nu = \pm 1$, $\pm 2$, and 4 as a function of $B$ ($\tau_{-}$ is discussed in Supplemental Material).

We first discuss the behavior of $\mu$, which corresponds to the time-of-flight and is related to the group velocity $v$ of EMPs as $\mu \propto 1/|v|$.
We find that $\mu$ takes discrete values, almost independent of $B$, for each $\nu$.
It is well established that $v$ of EMP modes screened by a metallic top gate electrode is given by
\begin{equation}
\label{VelocityEqn}
v = \frac{\sigma_{xy}}{\varepsilon} \frac{d}{w}
\equiv \frac{\sigma_{xy}}{C_{g}},
\end{equation}
where $\varepsilon$ is the dielectric constant, $d$ is the distance between the gate and the 2D system, $w$ is the transverse width of EMP modes, and $C_{g} \equiv \varepsilon w/d$ is the effective gate capacitance per unit length~\cite{kamata2010voltage, kumada2011edge, kumada2020suppression}.
Since $\sigma_{xy} \propto \nu$ in Eq.~(\ref{VelocityEqn}), we expect $\mu \propto 1/|\nu|$, which we confirm for both electrons and holes in Fig.~\ref{Fig4}(d), where $\mu$ averaged over different $B$ is plotted as a function of $1/\nu$.
From the slopes of the $\mu$ vs $1/\nu$ plot and the path length $L = 30$~$\mu$m, we obtain $v = (1.30 \pm 0.03) \times 10^{4} \cdot \nu$~m/s for electrons and $|v| = (1.43 \pm 0.04) \times 10^{4} \cdot |\nu|$~m/s for holes [Fig.~\ref{Fig4}(e)].

Equation~(\ref{VelocityEqn}), valid when $d \ll w$, tells us that the screening by the top gate reduces the EMP velocity by the factor $d/w$.
In the bulk region under the global gate, the distances $d$ between the gate and the centers of the InAs and In$_{\rm 0.25}$Ga$_{\rm 0.75}$Sb wells hosting the 2D electrons and holes are 140 and 148~nm, respectively.
Putting these together with the measured $|v|$ and effective dielectric constant of $\varepsilon = 12 \varepsilon_{0}$ (for details of the layer structure, see Appendix A), we estimate the transverse width $w$ of EMPs to be $\sim 4$~$\mu$m for both electrons and holes.
We note that this value is much larger than the expected width of edge channels ($\lesssim 100$~nm) defined by a sharp edge potential characteristic of InAs quantum wells~\cite{akiho2019counterflowing}.
It is worth noting that the value is even larger than the one, $0.8$~$\mu$m, for EMPs in a top-gated InAs/Al$_{\rm 0.7}$Ga$_{\rm 0.3}$Sb single quantum well estimated from the measured velocity~\cite{kumada2020suppression} with the exact sample structure taken into account.

One plausible origin of this unexpectedly large $w$ is charge puddles in the bulk.
Recently, it has been shown that, in the presence of a top gate electrode, transport properties of EMPs are strongly influenced by charge puddles in the bulk through capacitive coupling through the gate~\cite{lin2021time}.
Consequently, in the low-frequency limit, $C_{g}$ in Eq.~(\ref{VelocityEqn}) is replaced by $C_{g} + C_{p}$, where $C_{p}$ is the gate-puddle capacitance per unit length~\cite{lin2021time}, resulting in a significant reduction in the EMP velocity.
The influence of charge puddles will become more significant as the band gap narrows, which might explain the larger $w$ in the InAs/In$_{\rm 0.25}$Ga$_{\rm 0.75}$Sb CQW than in the InAs/Al$_{\rm 0.7}$Ga$_{\rm 0.3}$Sb single quantum well.
The puddle scenario is also consistent with the recent microwave spectroscopy on narrow-gap HgTe-based 2D topological insulators with a top gate, indicating unexpectedly large density of states from edge states~\cite{dartiailh2020dynamical}.

Coupling with charge puddles is expected to affect not only the mode velocity but also the waveform of the EMP pulse.
This is seen in the behavior of $\tau_{+}$, which reflects the broadening of EMP pulse.
As shown in Fig.~\ref{Fig4}(c), $\tau_{+}$ takes discrete values for each $\nu$, with no clear dependence on $B$.
We find that, when averaged over different $B$, $\tau_{+}$ increases simply in proportion to $\mu$ [Fig.~\ref{Fig4}(f)].
Since in the presence of screening by a gate electrode the dispersion of EMP modes becomes linear, with a wave-number-independent velocity [Eq.~(\ref{VelocityEqn})], the further broadening of the pulse with the time-of-flight cannot arise from dispersion.
We therefore suggest that the broadening is caused by the dissipation through the coupling to the puddles.

We note that, although the influence of charge puddles can be reduced by increasing the energy gap of the QH state with increasing $B$~\cite{lin2021time}, our data show no clear dependence on $B$ [Fig.~\ref{Fig4}(c)].
This may be related to the limited time-resolution due to the relatively short propagation length ($L = 30$~$\mu$m) as compared with that in Ref.~\cite{lin2021time} ($L = 420$~$\mu$m).
While further investigation is needed to clarify the precise mechanism of the influence from charge puddles in our device, the above results highlight the advantage of performing time-resolved measurements in CQWs, which allows us to access charge dynamics in QH chiral edge channels of both the electron and hole regimes.

\section{Summary}
\label{Summary}

We developed an on-chip time-resolved transport measurement scheme for QH chiral edge states in narrow-gap systems.
The ambipolar character of the InAs/In$_{x}$Ga$_{1-x}$Sb CQWs enables us to investigate nonequilibrium charge dynamics in both the electron and hole regimes separately using a single device.
Time-resolved pulsed charges are observed only for one field direction opposite for electrons and holes, reflecting the chirality of each carrier.
By analyzing the measured waveforms, we can discuss dissipative charge transport in the edge states.

Our time-resolved measurement scheme is applicable to other low-dimensional systems, including exfoliated graphene and 2D/3D topological materials, and not limited to QH chiral edge states.
Such experiments will pave the way for investigations of dynamical properties of topological edge states such as QSH helical edge states or chiral Majorana edge states.

\section*{Acknowledgments}
The authors thank H.~Murofushi and S.~Sasaki for sample fabrication and E.~Bocquillon for fruitful discussion.
This work was supported by JST, PRESTO Grant Number JPMJPR20L2, Japan.

\newpage

\section*{Appendix A: Device fabrication}
To fabricate the structure shown in Fig.~\ref{Fig1}(a), first, Ohmic contacts were fabricated by etching to the InAs layer prior to evaporation of Ti/Au (10/120~nm).
Then, the entire mesa defined by wet etching was covered with a 40-nm-thick Al$_{\rm 2}$O$_{\rm 3}$ gate dielectric by atomic layer deposition (ALD), followed by evaporation of Ti/Au (15/280~nm) for the fine gates.
Finally, the entire structure was covered with a 40-nm-thick Al$_{\rm 2}$O$_{\rm 3}$ gate dielectric by ALD, followed by evaporation of Ti/Au (15/320~nm) for the global gate in regions other than the injection and detection gates.
The global gate has slight overlaps with the fine gates to eliminate regions with uncontrolled densities while keeping the capacitive coupling to a minimum.
With the 80-nm-thick Al$_{\rm 2}$O$_{\rm 3}$ layer, the distances from the global gate to the 2D electrons and holes are 140 and 148~nm, respectively.
The effective dielectric constant of $\varepsilon = 12 \varepsilon_{0}$ is the averaged value of the Al$_{\rm 2}$O$_{\rm 3}$ layer ($9 \varepsilon_{0}$) and the CQW ($15 \varepsilon_{0}$).

\section*{Appendix B: RF measurement sequence}
The injection pulse $V_{\rm inj} (t)$ with a repetition frequency of 25~MHz was modulated with a 1-MHz sine wave on the AWG.
The time delay $t_{d}$ was also controlled on the AWG with the time resolution of 1/(64~GSa/s) $\sim$ 15~ps.
The current $I (t_{d})$ was amplified by a room-temperature amplifier and filtered by a 1-MHz band-pass filter.
The data recorded on the digitizer (100,000~traces taken with the duration of 10~$\mu$s) were numerically averaged and then demodulated with a 1-MHz sine wave to extract target signals in the gigahertz range.

The amplitude modulation (AM) technique, which is commonly used in electronic communication, enables us to carry target high-frequency signals via a much lower frequency wave. As a result, undesirable high-frequency crosstalk and dissipation can be suppressed with this AM measurement scheme.

\newpage
\renewcommand{\thefigure}{S\arabic{figure}}
\setcounter{figure}{0}
\section*{Supplemental Material}

\noindent{\bf 1. Magnetotransport measurements in a Hall-bar sample}
\begin{figure}[h]
\begin{center}
\includegraphics[scale=1.5]{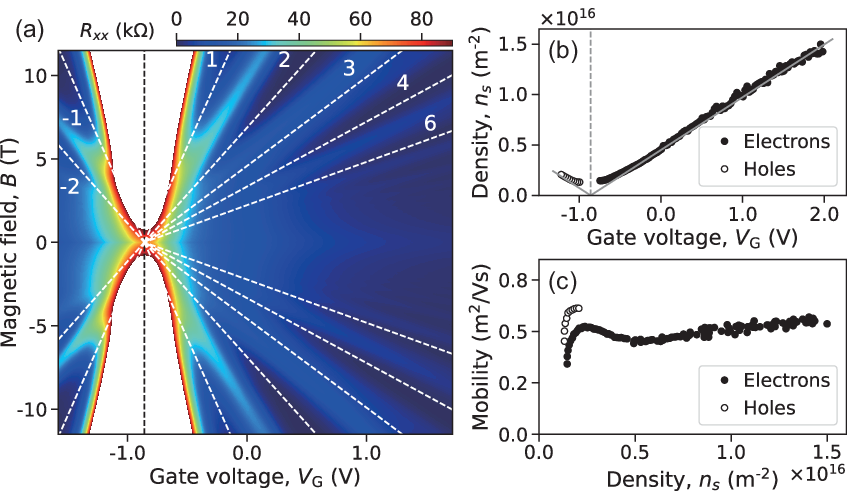}
\caption{
(a)
Color-scale plot of $R_{xx}$ as a function of $V_{\rm G}$ and $B$.
The dashed white lines represent the positions where the filling factors take integer values.
The vertical dashed black line represents the CNP.
(b)
Carrier density $n_{s}$ of electrons (solid circles) and holes (solid circles) as a function of $V_{\rm G}$.
(c)
Mobility of electrons (solid circles) and holes (solid circles) as a function of $n_{s}$.
}
\label{FigS1}
\end{center}
\end{figure}

Figure~\ref{FigS1}(a) shows the longitudinal resistance $R_{xx}$ as a function of top-gate voltage $V_{\rm G}$ and magnetic field $B$ measured at 1.5~K in a standard Hall-bar sample (50~$\mu$m in width and 150~$\mu$m in length) fabricated from the same composite quantum well used in the main text.
The large resistance peak at $V_{\rm G} = -0.86$~V (dashed black line) corresponds to the charge neutrality point (CNP), where each carrier density $n_{s}$ is zero [Fig.~\ref{FigS1}(b)].
At high fields, the $R_{xx}$ shows the quantum Hall effect without a signature of an inverted band structure.
Figure~\ref{FigS1}(c) shows each carrier mobility as a function of $n_{s}$.
As $n_{s}$ is increased, the electron and hole mobilities increase and reach 0.50 and 0.61~m$^{2}$/Vs at $n_{s} = 2.0 \times 10^{15}$~m$^{-2}$, respectively.

\newpage

\noindent{\bf 2. Detection condition $\nu^{\prime} = \nu$}
\begin{figure}[h]
\begin{center}
\includegraphics[scale=1.5]{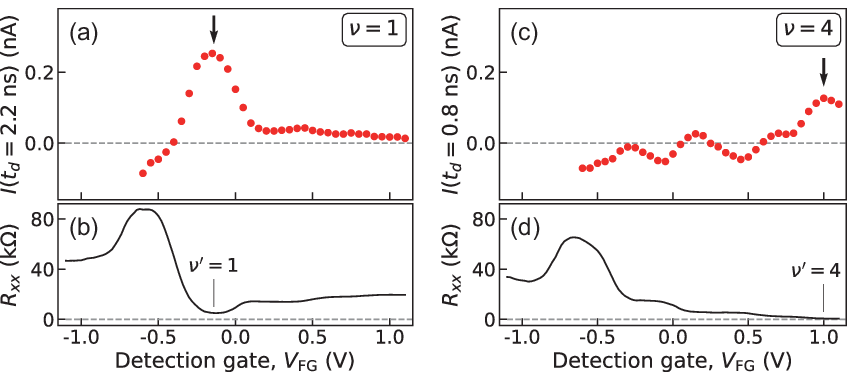}
\caption{
(a) $I(t_{d} = 2.2$~ns) and (b) $R_{xx}$ for $\nu = 1$ as a function of $V_{\rm FG}$. 
Panels (c) and (d) represent the same plots for $\nu = 4$, where the current peak appears at $t_{d} = 0.8$~ns.
Arrows represent the positions $\nu^{\prime} = \nu$.
}
\label{FigS2}
\end{center}
\end{figure}

As shown in Fig.~2(b) of the main text, at the filling factor $\nu = 2$, the measured signal amplitude takes a maximum value at $\nu^{\prime} = \nu = 2$.
This detection condition $\nu^{\prime} = \nu$ is suitable for other filling factors.
Figures~\ref{FigS2}(a) and \ref{FigS2}(c) show the current peak value for $\nu = 1$ and 4, respectively, as a function of $V_{\rm FG}$.
The signal amplitude becomes maximum at $\nu^{\prime} = \nu$ as indicated by arrows.

\noindent{\bf 3. $V_{\rm GG}$ dependence of the waveform at $B = 10$~T}
\begin{figure}[h]
\begin{center}
\includegraphics[scale=1.25]{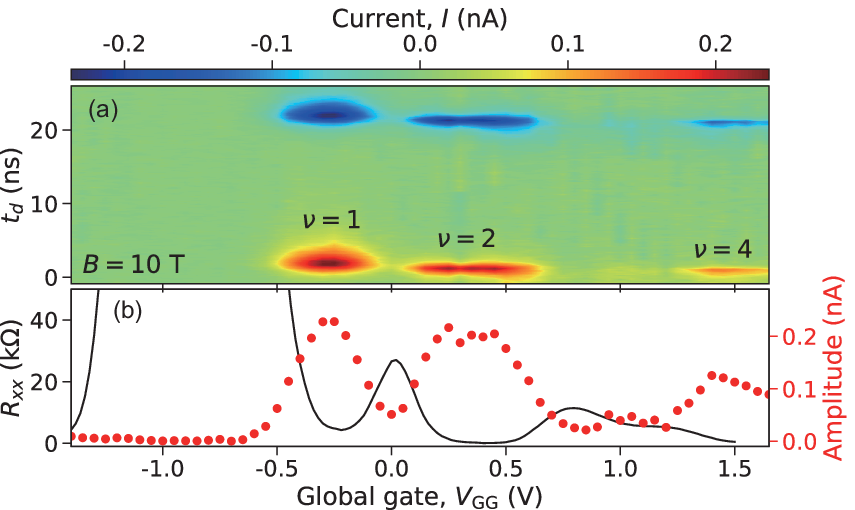}
\caption{
(a) Color-scale plots of $I$ as a function of $t_{d}$ and $V_{\rm GG}$ at $B = 10$~T.
(b) $R_{xx}$ (black line) and signal amplitude (red dots) as a function of $V_{\rm GG}$ at $B = 10$~T.
}
\label{FigS3}
\end{center}
\end{figure}

Figure~\ref{FigS3}(a) shows $I$ as a function of $t_{d}$ and $V_{\rm GG}$ measured at $B = 10$~T.
In the whole regime, positive and negative pulses have the same waveform with reverse polarity.
The peak shifts to larger $t_{d}$ as $V_{\rm GG}$ decreases (and hence $\nu$ decreases) and, at the same time, broadens along the time axis.
The signal amplitude takes maximum values in the quantum Hall regime [Fig.~\ref{FigS3}(b)].

\noindent{\bf 4. $B$ dependence of the waveform at integer $\nu$}
\begin{figure}[h]
\begin{center}
\includegraphics[scale=1.24]{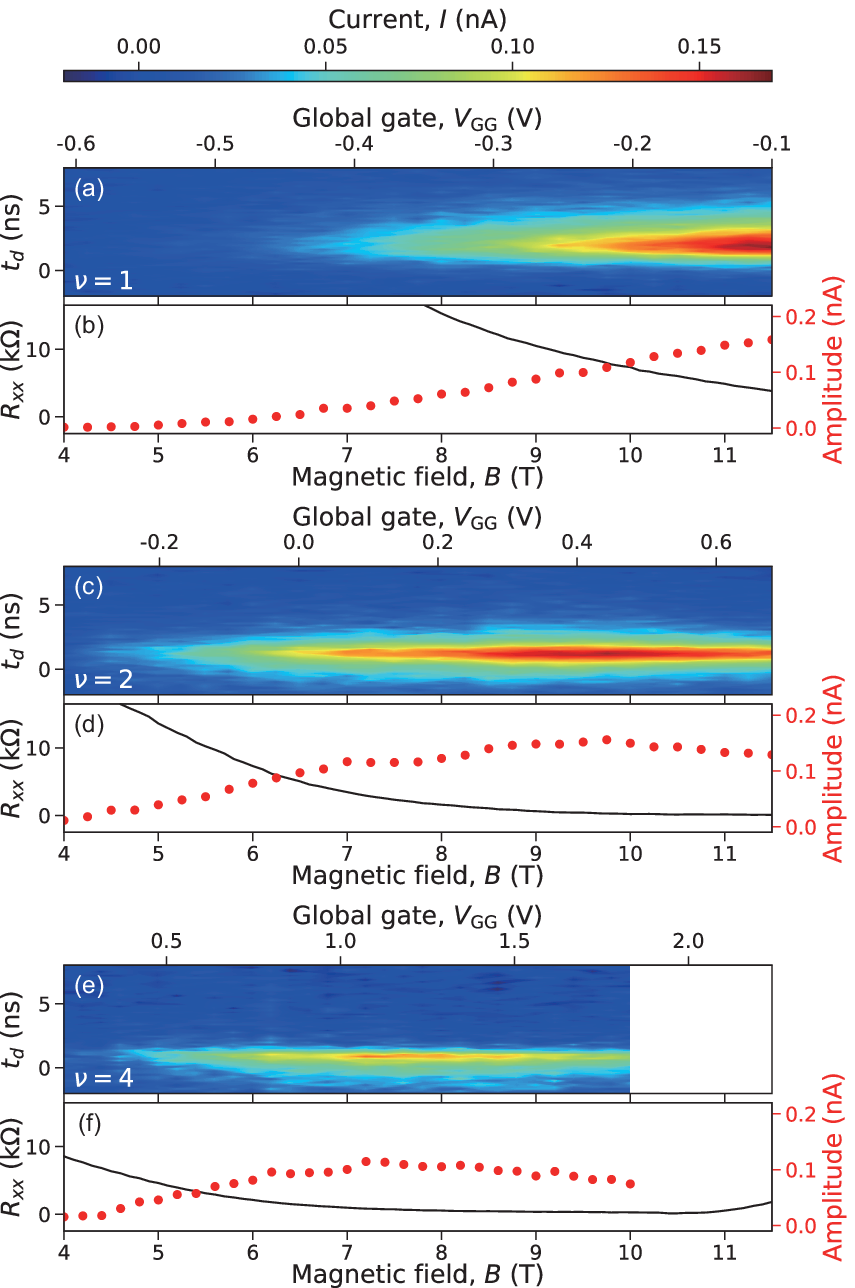}
\caption{
(a)
Color-scale plot of $I$ as a function of $t_{d}$ and $B$ for $\nu = 1$.
(b)
$R_{xx}$ (black line) and signal amplitude (red dots) as a function of $B$ for $\nu = 1$.
Panels (c), (d) and (e), (f) represent the same plots for $\nu = 2$ and $4$, respectively.
}
\label{FigS4}
\end{center}
\end{figure}

Figures~\ref{FigS4}(a), \ref{FigS4}(c), and \ref{FigS4}(e) show $I$ as a function of $t_{d}$ and $B$ measured at $\nu = 1$, $2$, and $4$, respectively, where carrier densities are simultaneously tuned with $V_{\rm GG}$ to maintain $\nu$ constant.
The waveform is almost $B$ independent, while the amplitude increases with decreasing $R_{xx}$ [Figs.~\ref{FigS4}(b), \ref{FigS4}(d), and \ref{FigS4}(f)].

\noindent{\bf 5. Waveform analysis}
\begin{figure}[h]
\begin{center}
\includegraphics[scale=1.5]{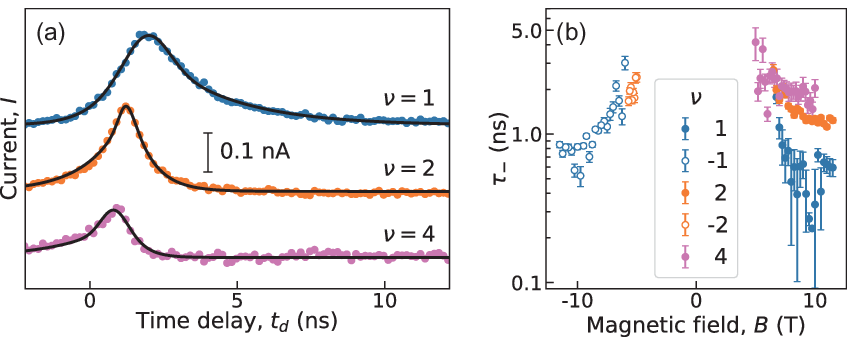}
\caption{
(a)
Waveforms for $\nu = 1$, 2 and $4$ measured at $B = 10$~T (dots) with the fitting curves $f_{+}(t - \mu) + f_{-}(t - \mu)$ (solid lines).
(b)
$B$ dependence of $\tau_{-}$ at each $\nu$.
}
\label{FigS5}
\end{center}
\end{figure}

Figures~\ref{FigS5}(a) shows $I(t_{d})$ traces for $\nu = 1$, 2 and 4 (dots) taken at $B = 10$~T with the fitting curves $f_{+}(t - \mu) + f_{-}(t - \mu)$ (solid lines).
$\tau_{-}$ obtained by the fitting corresponds to the exponential relaxation time on the left side of the waveforms and predominately reflects the response time of the local potential to the detection pulse~\cite{kamata2009correlation}. 
As shown in Fig.~\ref{FigS5}(b), $\tau_{-}$ increases with decreasing $B$ for all $\nu$.
We suggest that charge puddles influence the response time by increasing the capacitance to the detection gate for smaller $B$.
Note that the actual response time can be estimated by measuring and analyzing the waveform for plasmon pulses with various widths injected from an Ohmic contact as demonstrated in Ref.~\cite{kamata2009correlation}.

\newpage

%


\end{document}